\documentclass[epj]{svjour}

\usepackage{amsmath}
\usepackage{amsfonts}
\usepackage{amssymb}
\usepackage{graphics}
\usepackage{epsfig}

\begin{document}

\title{Construction of inflationary scenarios with the Gauss-Bonnet term and nonminimal coupling}

\author{Ekaterina~O.~Pozdeeva\thanks{\emph{e-mail:}  pozdeeva@www-hep.sinp.msu.ru} \and Sergey~Yu.~Vernov\thanks{\emph{e-mail:} svernov@theory.sinp.msu.ru}}
\titlerunning{ }
\authorrunning{ }
\institute{Skobeltsyn Institute of Nuclear Physics, Lomonosov Moscow State University, Leninskie Gory~1, 119991, Moscow, Russia}
\date{ {\ } }
\abstract{Inflationary models with a scalar field nonminimally coupled both with the Ricci scalar and with the Gauss-Bonnet term are studied. We propose the way of generalization of inflationary scenarios with the Gauss-Bonnet term and a scalar field minimally coupled with the Ricci scalar to the corresponding scenarios with a scalar field nonminimally coupled with the Ricci scalar. Using the effective potential, we construct a set of models with the same values of the scalar spectral index $n_s$ and the amplitude of the scalar perturbations $A_s$ and different values of the tensor-to-scalar ratio $r$.
 % end of PACS codes
} %end of abstract
\maketitle

\section{Introduction}

Cosmic inflation is a stage of an accelerated expansion of the early Universe evolution that provides a simple explanation of both the large scale structure we observe today and the fact that the Universe is approximately isotropic, homogeneous, and spatially flat at cosmological distances~\cite{Starobinsky:1980te,Sato:1980yn,Guth:1981uk,Linde:1981mu,Starobinsky:1982,Mukhanov:1982nu,Starobinsky:1983,ellis,SUSEinflation_0,Starobinsky:1986fx}. Inflationary models yield accurate quantitative predictions for observable quantities known as inflationary parameters. The current observational constraints on the inflationary parameters~\cite{Planck2018} give the values of the scalar spectral index $n_s$ and the amplitude of scalar perturbations $A_s$, whereas the tensor-to-scalar ratio $r$ is restricted only from above. These constraints show that the single-field inflationary models are realistic, but the simplest inflationary models with minimally coupled scalar fields should be ruled out.

Scalar fields (inflatons) play a central role in the current description of the evolution of the Universe at an early epoch~\cite{Linde:1981mu,Salopek:1990jq,Lyth:1998xn,Finelli:2008zg,Mukhanov:2013tua}.
The recent observation data support taking into account quantum properties of the inflaton because quantum corrections to the action of the scalar field minimally coupled to gravity include non-minimally coupling term~\cite{ChernikovTagirov,Tagirov,Callan:1970ze,BirrellDavies}.
An important step towards the unification of physics at all energy scales could be the possibility to describe the inflation using particle physics models. For this reason,
inflationary models with scalar fields connected with particle physics are actively investigated. We can mention models with the Standard Model Higgs boson as an inflaton~\cite{Cervantes-Cota1995,higgsinf_0,higgsinf_1,higgsinf_2,higgsinf_3,higssinflRG_0,higssinflRG_1,higssinflRG_2,KaiserHiggs,Lerner,Ren}, supersymmetric inflationary models~\cite{ellis,SUSEinflation_0,SUSEinflation_1,MazumdarRev,einhorn,Ferrara:2010yw,SUSEinflation_2,Ketov:2012jt,Pallis:2013yda,Dubinin:2017irg,DubininMSSM2,Ketov:2021fww}, and inflationary scenarios connected with nonsupersymmetric grand unified theories~\cite{Starobinsky:1982,nonmin-quant,GUT_Inflation,GUT_Inflation1,GUT_Inflation2,OdintsovNOnmin,EOPV2014,Elizalde:2015nya,Pozdeeva:2016hrz}. Most of these models include a nonminimal coupling between scalar fields and the Ricci scalar.

The standard way to analyze such inflationary models includes the conformal transformation of the metric to formulate them as equivalent models with minimally coupled scalar fields. It has been shown in Ref.~\cite{Kaiser,Kaiser2} that the inflationary parameters can be considered as invariants under this metric transformation with good accuracy. In other words, in the slow-roll approximation, the observable inflationary parameters are the same in both frames. Moreover, the frame-independent classification of single-field inflationary models has been proposed using the expressions of the slow-roll parameters and the relevant observables in terms of frame invariant quantities~\cite{Jarv:2016sow}. $F(R)$ gravity models can be reformulated as the General Relativity models with scalar fields as well. For example, the $R^2$-inflation~\cite{Starobinsky:1980te,Starobinsky:1983,Vilenkin:1985md,Mijic:1986iv,Maeda1988} and the Higgs-driven inflation~\cite{higgsinf_0} have almost the same values of inflationary parameters~\cite{BezrukovR2Higgs}.
The simple reason why the Starobinsky $R^2$ inflation and the Higgs-driven inflation produce the same predictions for the inflationary parameters $n_s$ and $r$  is the possibility to omit in the slow-roll approximation the Higgs gradient term in the Jordan frame action~\cite{Elizalde:2015nya,He:2018gyf}. Note that each of these models includes only one arbitrary parameter.

In this paper, we consider the gravity model, described by the following action:
\begin{equation}
\label{action1}
S=\int d^4x\frac{\sqrt{-g}}{2}\left[F(\phi)R-g^{\mu\nu}\partial_\mu\phi\partial_\nu\phi-2V(\phi)-\xi(\phi){\cal G}\right],
\end{equation}
where the functions $F(\phi)$, $V(\phi)$, and $\xi(\phi)$ are differentiable ones,  $R$ is the Ricci scalar and
\begin{equation*}
\mathcal{G}=R_{\mu\nu\rho\sigma}R^{\mu\nu\rho\sigma}-4R_{\mu\nu}R^{\mu\nu}+R^2
\end{equation*}
is the Gauss-Bonnet term. We assume that $F(\phi)>0$ and $V(\phi)>0$ during inflation.

There are a few reasons to add the Gauss-Bon\-net term multiplied to a function of a scalar field in the action. On the one hand, this term arises in the string theory framework as a quantum correction to the Einstein-Hilbert action~\cite{Cartier:2001is,Hwang:2005hb,Sami:2005zc,Tsujikawa:2006ph}. On the other hand, for inflationary scenarios with a nonminimal coupling between a scalar field and the Ricci term, one usually assumes that after inflation the scalar field tends to a constant and the model tends to the General Relativity one. Considering inflationary models in the framework of the Einstein-Gauss-Bonnet gravity, it is natural to assume that a scalar field non-minimally coupled not only with the Ricci scalar but also with the Gauss-Bonnet term. For example, the Higgs-driven inflation with the Gauss-Bonnet term has been considered in Refs.~\cite{vandeBruck:2015gjd,JoseMathew}. In such a way, one gets another model with the same behaviour at a late time, but with different evolution during inflation.

There are a lot of inflationary scenarios in Ein\-stein-Gauss-Bonnet gravity~\cite{vandeBruck:2015gjd,JoseMathew,Guo,Guo10,Guo13,Koh,Koh18,Hikmawan:2015rze,Koh:2016abf,Nozari:2017rta,Yi:2018gse,Chakraborty:2018scm,Odintsov:2018zhw,Fomin:2019yls,Kleidis:2019ywv,Rashidi:2020wwg,Odintsov:2020sqy,Odintsov:2020sqz,Fomin:2020hfh,m,Pozdeeva:2020apf,Gao:2020cvb,Oikonomou:2020tct,Odintsov:2021lum,Venikoudis:2021irr,Pozdeeva:2021nmz}. The most of them includes a constant function~$F$~\cite{Guo,Guo10,Guo13,Koh,Koh18,Hikmawan:2015rze,Koh:2016abf,Yi:2018gse,Odintsov:2018zhw,Kleidis:2019ywv,Rashidi:2020wwg,Odintsov:2020sqy,Odintsov:2020sqz,Fomin:2020hfh,m,Pozdeeva:2020apf,Gao:2020cvb,Pozdeeva:2021nmz}.
Note that for models with a constant $F$, the problem of reconstructing the function $V$ and $\xi$ from the observational data has been considered in Ref.~\cite{Koh:2016abf}.
The goal of this paper is to generalize such inflationary scenarios to the case of a positive function~$F$. We analyze the possibility to construct a set of inflationary models with the same values of the scalar spectral index~$n_s$ and the amplitude of the scalar perturbation~$A_s$ starting from the known model with a scalar field minimally coupled with the Ricci scalar.  There is an important difference between inflationary models with and without the Gauss-Bonnet term. A conformal transformation of the metric is not a useful tool for the investigation of inflationary models with the Gauss-Bonnet term. For this reason, it is not obvious how the knowledge of a suitable Einstein-Gauss-Bonnet gravity inflationary scenario with minimal coupling can assist in the construction of inflationary scenarios with nonminimal coupling. In our paper, we clarify this question.

We formulate a new method for the construction of appropriate slow-roll inflationary models due to the effective potential method. Such a method allows us to reproduce spectral index, the amplitude of scalar perturbation, and we should only check the tensor-to-scalar ratio for considering models with nonminimal coupling. A standard way of the reconstruction of inflationary models~\cite{Mukhanov:2013tua,Koh:2016abf,m} includes the assumption of an explicit dependence of the inflationary parameter~$n_s$ and~$r$ as functions of the e-folding number~$N_e$. A broad class of inflationary scenarios with different nonminimal coupling between the Ricci scalar and the inflaton predicts the same functions $n_s(N_e)$ and $r(N_e)$. This fact is actively used to construct inflationary scenario in the cosmological attractor approach~\cite{Kallosh:2013hoa,Roest:2013fha,LindeKallosh,Kallosh:2014rha,LindeKallosh2,Binetruy:2014zya,LindeKallosh1,Carrasco:2015pla,Ventury2015,Pieroni2015,Elizalde:2015nya,DubininMSSM2,Nozari:2017rta,Odintsov:2020thl,Rodrigues:2021olg}. The generalization of this approach to models with the Gauss-Bonnet term has been made in~\cite{Koh:2016abf,Nozari:2017rta,m}. It has been shown in Ref.~\cite{Pozdeeva:2020apf} that $n_s(N_e)$ can be expressed via the effective potential proposed in Ref.~\cite{Pozdeeva:2019agu}  (see also, Ref.~\cite{Vernov:2021hxo}) for models with the Gauss-Bonnet term.\footnote{The effective potential for the model with nonminimally coupled scalar field without the Gauss-Bonnet term has been proposed and used in Refs.~\cite{Skugoreva:2014gka,Pozdeeva:2016cja,Jarv:2021qpp}.} In this paper, we show that the effective potential is a useful tool that allows us to generalize the known inflationary models with the Gauss-Bonnet term. In distinguish to the cosmological attractor approach, we construct inflationary models with the same functions $n_s(N_e)$ and $A_s(N_e)$, but with different functions $r(N_e)$.

The paper is organized as follows. In Section 2, we remind the main formulae about the slow-roll regime in Einstein-Gauss-Bonnet gravity and express the scalar spectral index $n_s$ and the amplitude of the scalar perturbation $A_s$ via the effective potential.  In Section 3, we show how the effective potential can be used to check the possibility of construction of an inflationary model with the given functions $n_s(N_e)$ and $r(N_e)$. In Section 4, we propose the way of generalization of the Gauss-Bonnet inflationary model with a scalar field minimally coupled with the Ricci scalar and construct inflationary models with nonminimal couplings. Section~5 is devoted to our conclusions.

\section{Slow-roll regime in Einstein-Gauss-Bonnet gravity with nonminimal coupling}

Let us consider the Einstein-Gauss-Bonnet gravity model, described by action~(\ref{action1}).
In the spatially flat Friedmann-Lema\^{i}tre-Robert\-son-Walker metric with
\begin{equation*}
ds^2={}-dt^2+a^2(t)\left(dx^2+dy^2+dz^2\right),
\end{equation*}
one obtains the following system of evolution equations~\cite{vandeBruck:2015gjd,Pozdeeva:2019agu}:
 \begin{eqnarray}
              && 6H^2\left(F-4H\xi_{,\phi}\dot{\phi}\right) = \dot{\phi}^2+2V-6HF_{,\phi}\dot{\phi}, \label{equ00} \\
              &&2\dot{H}\left(F-4H\xi_{,\phi}\dot{\phi}\right)=\nonumber \\
              &{}&4H^2\left(\ddot{\xi}-{\dot{\phi}}^2-H\xi_{,\phi}\dot{\phi}\right) -\ddot{F}+HF_{,\phi}\dot{\phi}, \label{equ11}\\
               && \ddot{\phi}+3H\dot{\phi}=\nonumber\\
                &{}&3\left(\dot{H}+2H^2\right)F_{,\phi}-V_{,\phi} -12\xi_{,\phi} H^2\left(\dot{H}+H^2\right),\label{equphi}
\end{eqnarray}
where  $H=\dot{a}/a$ is the Hubble parameter, $a(t)$ is the scale factor, dots denote the derivatives with respect to the cosmic time  $t$ and
$A_{,\phi}\equiv {dA}/{d\phi}$ for any function $A(\phi)$.

 In the slow-roll approximation, defined by the following conditions~\cite{vandeBruck:2015gjd}:
 \begin{eqnarray}
 &&\dot{\phi}^2\ll V, \qquad |\ddot{\phi}|\ll 3H|\dot{\phi}|, \nonumber\\
 &&4|\dot{\xi}|H\ll F,\qquad |\ddot{\xi}|\ll H |\dot{\xi}|, \label{slow-roll-equa}\\
 && |\ddot{F}|\ll H|\dot{F}|\ll H^2F\nonumber,
 \end{eqnarray}
Eqs.~(\ref{equ00})--(\ref{equphi}) are:
\begin{eqnarray}
&&3FH^2\simeq {V},  \quad\quad\quad\quad\quad\label{equ00SLR} \\
&&2F\dot{H}\simeq{}-{\dot{\phi}^2}-4H^3\xi_{,\phi}\dot{\phi}+HF_{,\phi}\dot{\phi},\label{equ11SLR}  \quad\quad\quad\quad\quad\\
&&\dot{\phi}\simeq{}-\frac{V_{,\phi} +12\xi_{,\phi} H^4-6H^2F_{,\phi} }{3H}\label{dot(phi)}. \quad\quad\quad \quad\quad
\end{eqnarray}

To describe the Universe evolution during inflation we use the dimensionless parameter $N_{e}=-\ln(a/a_{e})$ as a new measure of time.
The constant $a_e$ is fixed by the condition that the end of inflation happens at $N_e=0$. The function $H(t)$ is always finite and positive during inflation, hence, $N_e$ is a monotonically decreasing function. The parameter $N_e$ has been used in Refs.~\cite{Mukhanov:2013tua,m}. Note that in many papers~\cite{Finelli:2008zg,vandeBruck:2015gjd,Guo,Odintsov:2020sqy,Pozdeeva:2020apf} the parameter $N=-N_e$ is used as a new independent variable for evolution equations\footnote{The advantage of using the e-folding number $N$ instead of cosmic time $t$ was shown in the context of the stochastic approach to inflation in Refs.~\cite{Starobinsky:1986fx,Finelli:2008zg}.}.

From Eqs.~(\ref{equ00SLR})--(\ref{dot(phi)}), we get the following leading-order equations:
\begin{eqnarray}
   &&\ln(H)^\prime=2W_{,\phi} V_{eff,\phi}\,, \label{EquHloN}\\
    &&  {\phi}^\prime=4WV_{eff,\phi}\,, \label{phiPrime}
\end{eqnarray}
where derivatives with respect to $N_e$ are denoted by pri\-mes, $W\equiv V/F$ and the effective potential~\cite{Pozdeeva:2020apf,Pozdeeva:2019agu,Vernov:2021hxo}:
\begin{equation}
\label{Veff}
V_{eff}(\phi)=\frac{1}{3}\xi(\phi)-\frac{F^2(\phi)}{4V(\phi)}.
\end{equation}

The  slow-roll approximation (\ref{slow-roll-equa}) requires
\begin{equation*}
|\epsilon_i|\ll1, \quad |\delta_i|\ll1, \quad |\zeta_i|\ll1,
\end{equation*}
where the slow-roll parameters are as follows~\cite{vandeBruck:2015gjd}:
\begin{eqnarray}
  &&\epsilon_1=\frac{(H^2)^\prime}{2H^2}\simeq\frac{W^\prime}{2W},\quad \epsilon_{i+1}={}- \frac{\epsilon_i^\prime}{\epsilon_i},\quad i\geqslant 1, \label{epsiloni}\\
 && \zeta_1={}-\frac{F^\prime}{F}, \quad \zeta_{i+1}={}- \frac{ \zeta_i^\prime}{ \zeta_i},\quad i\geqslant 1, \label{zetai}\\
 && \delta_1= {}-\frac{4H^2\xi^\prime}{F}\simeq{}-\frac{4V\xi^\prime}{3F^2},\quad \delta_{i+1}={}- \frac{ \delta_i^\prime}{ \delta_i},\, i\geqslant 1.\, \label{deltai}
\end{eqnarray}

It is easy to get:
\begin{equation*}
%\label{slrparam2}
\epsilon_2=2\epsilon_1-\frac{W^{\prime\prime}}{W^\prime},\,
 \zeta_2=-\zeta_1-\frac{F^{\prime\prime}}{F^\prime}, \,
\delta_2=-2\epsilon_1-\zeta_1-\frac{\xi^{\prime\prime}}{\xi^\prime}.
\end{equation*}

The tensor-to-scalar ratio $r$ is presented in terms of the slow-roll parameters as follows~\cite{vandeBruck:2015gjd}:
\begin{equation}
  r=8|2\epsilon_1+\zeta_1-\delta_1|=8\left|\frac{\left(H^2\right)^\prime}{H^2}-\frac{F^\prime}{F}+\frac{4H^2\xi^\prime}{F}\right|
  \label{r_i}.
\end{equation}
Using the derivative of the effective potential
\begin{equation}
\label{VeffN1}
    V_{eff}^\prime=\frac{F^2}{4V}\left(\frac{V'}{V}-2\frac{F'}{F}+\frac{4V}{3F^2}\xi'\right)=\frac{F^2}{32V}r,
\end{equation}
and Eq.~\eqref{phiPrime}, we obtain
\begin{equation}
{\phi^\prime}^2=4WV^\prime_{eff}.\label{phiPrime2N}
 \end{equation}
 
From here, we get the following relation between the tensor-to-scalar ratio $r$ and a square of the field derivative:
\begin{equation}\label{rphiN1}
    r=\frac{32W}{F}V_{eff}^\prime=\frac{8}{F}\left(\phi^\prime\right)^2.
\end{equation}

The spectral index of scalar perturbations $n_s$ has the following form:
\begin{equation}
  n_s=1-2\epsilon_1-\zeta_1-\frac{2\epsilon_1\epsilon_2+\zeta_1\zeta_2-\delta_1\delta_2}{2\epsilon_1+\zeta_1-\delta_1}
  =1-2\epsilon_1-\zeta_1+\frac{r^\prime}{r},\label{n_s_i}
\end{equation}
where we simplify the expression for $n_s$ using Eq.~(\ref{r_i}).
Introducing parameters
\begin{equation*}
\eta_0=\frac{H^2}{F}=\frac{V}{3F^2},\quad\eta_1=\frac{\eta_0^\prime}{\eta_0}=2\epsilon_1+\zeta_1,
\end{equation*}
we present $n_s$ via derivatives of the effective potential:
\begin{equation}
\label{nsN}
  n_s=1+\frac{d\ln\left({r}/{\eta_0}\right)}{dN}=1+\frac{d\ln\left({F^2r}/{V}\right)}{dN}=1+\frac{V_{eff}^{\prime\prime}}{V_{eff}^\prime}.
\end{equation}
Note that this formula generalizes the result obtained  in the case of a constant $F$ in Ref.~\cite{Pozdeeva:2020apf}.
During inflation $V_{eff}^\prime=4W(V_{eff,\phi})^2>0$, so the condition $V_{eff}^{\prime\prime}<0$ is necessary to get $n_s<1$.

The expression of the amplitude of the scalar perturbations in terms of the effective potential is  as follows:
\begin{equation}
A_s\simeq\frac{2H^2}{\pi^2 F r}\simeq\frac{2W}{3\pi^2 F r}=\frac{1}{48\pi^2 V^\prime_{eff}},\label{As}
\end{equation}
where we have used Eq.~(\ref{VeffN1}) to get the last equality.

To reconstruct an inflationary model we assume an explicit form of the inflationary parameters as functions of $N_e$. Formula (\ref{nsN}) shows how the knowledge of $n_s(N_e)$ allows to calculate $V_{eff}^\prime(N_e)$~\cite{Mukhanov:2013tua,Koh:2016abf,m}.  The inflationary parameters  $n_s$ and $A_s$ are defined by $V_{eff}^\prime(N_e)$ only. So, if we assume a some form of $n_s(N_e)$, then integrating Eq.~(\ref{nsN}), we get
\begin{equation}
\label{DVeffN}
{V_{eff}^\prime}(N_e)=\exp\left(\int\limits_{N_1}^{N_e} (n_s(\tilde{N})-1)d\tilde{N}\right),
\end{equation}
where $N_1$ is an arbitrary constant, that can be fixed after substituting $V_{eff}^\prime(N_e)$ into Eq.~(\ref{As}).

\section{Inflationary scenarios with the given functions $n_s(N_e)$ and $r(N_e)$}

Let us try to construct models with the Gauss-Bonnet term in which the functions $n_s(N_e)$ and $r(N_e)$ are coincide in the leading order of $1/N_e$ with conformal attractor models~\cite{Kallosh:2013hoa}:
\begin{equation}
n_s=1-\frac{2}{N_{e}+N_0}\,,
\label{ns_cat}
\end{equation}
\begin{equation}
\label{r_cat}
    r=\frac{12C_\alpha}{(N_{e}+N_0)^2},
\end{equation}
where $|N_0|\ll 60$ and $C_\alpha>0$ are constants. The Starobinsky $R^2$ inflation~\cite{Starobinsky:1980te}
and the Higgs-driven inflation~\cite{higgsinf_0} correspond to $C_{\alpha}=1$.

Substituting (\ref{ns_cat}) into Eq.~(\ref{nsN}),  we obtain:
\begin{equation}
\label{VeffN}
V_{eff}^\prime(N_e)=C_{eff}(N_e+N_0)^{-2}=\frac{C_{eff}}{4}(n_s-1)^2,
\end{equation}
where $C_{eff}>0$ is an integration constant.
Using Eq.~(\ref{As}), we get
\begin{equation}
\label{Aest}
A_s=\frac{1}{12\pi^2 C_{eff}(n_s-1)^2}.
\end{equation}

From Eq.~(\ref{n_s_i}), it follows that $\eta_1=0$ for $n_s$ and $r$ are given by (\ref{ns_cat}) and  (\ref{r_cat}). So, in this case $V=CF^2$, where $C$ is a positive constant. If $F$ is a constant, we obtain from Eq.~(\ref{VeffN1}) that the potential $V$ is a constant as well and the slow-roll approximation does not work, because Eq.~(\ref{dot(phi)}) gives that $H$ is a constant and $\epsilon_1=0$. For an arbitrary nonconstant $F(\phi)$, we obtain that $\zeta_1=-2\epsilon_1$ and during inflation $\zeta_1<-1$. Such inflationary scenarios can be possible but correct calculation of the inflationary parameters should include a numerical integration of the Eqs.~(\ref{equ00})--(\ref{equphi}) without any approximation. We do not consider such inflationary scenarios and use another way for constructing of inflationary scenarios with a nonconstant
function~$F$.

\section{Generalization of inflationary models with minimal coupling}

\subsection{The model with an exponential potential $V$ and a constant function $F$}

The values of the inflationary parameters $n_s$ and $r$ given by (\ref{ns_cat}) and (\ref{r_cat}) are not suitable.
By this reason, we assume that $n_s$ described by Eq. \eqref{ns_cat} only in large $N_e$ approximation and has a more complicated dependence on~$N_e$:
\begin{equation}
n_s=1-\frac{2}{N_e+N_0}+\frac{C_2}{(N_e+N_0)^2},
\label{ns_C2}
\end{equation}
where a constant $|C_{2}|\ll 60$. The corresponding effective potential has the following form:
\begin{equation}\label{Veffexp}
    V_{eff} = C_{eff}\exp\left(-\frac{C_2}{N_e+N_0}\right).
\end{equation}
From Eq.~(\ref{As}), we obtain
\begin{equation}\label{As2}
    A_s=\frac{(N_e+N_0)^2}{48\pi^2C_{eff}C_2}\exp\left(\frac{C_2}{N_e+N_0}\right)\,.
\end{equation}

In Ref.~\cite{m}, inflationary models with the Gauss-Bonnet term and a constant $F$ have been constructed by using explicit functions $n_s(N_e)$ and $r(N_e)$ given by (\ref{ns_C2}) and  (\ref{r_cat}) correspondingly.
In this model, the function $F=M_{Pl}^2$, the potential
\begin{equation}
\label{Vmincoupling}
\tilde{V}=V_0\exp\left({}-\omega_0\exp\left(-\sqrt{\frac{2}{3C_\alpha}}\frac{\phi}{M_{Pl}}\right)\right),
\end{equation}
and
\begin{equation}
\label{ximincoupling}
\tilde{\xi}=\xi_0\exp\left(\omega_0\exp\left(-\sqrt{\frac{2}{3C_\alpha}}\frac{\phi}{M_{Pl}}\right)\right),
\end{equation}
where $V_0>0$, $C_\alpha>0$, $\omega_0$, and $\xi_0$ are constants.
The effective potential is
\begin{equation}
\label{VeffC2}
\tilde{V}_{eff}=\frac{4\xi_0V_0-3M_{Pl}^4}{12V_0}\exp\left(\omega_0\exp\left(-\sqrt{\frac{2}{3C_\alpha}}\frac{\phi}{M_{Pl}}\right)\right).
\end{equation}

Using Eq.~(\ref{phiPrime}), we obtain $\phi(N_e)$ in the slow-roll approximation:
\begin{equation}
\label{phiN2}
    \phi(N_e)=\frac{\sqrt{6C_\alpha}}{2}M_{Pl}\ln\left(\frac{2\omega_0(3M_{Pl}^4-4V_0\xi_0)}{9C_\alpha M_{Pl}^4}(N_e+N_0)\right),
\end{equation}
where $N_0$ is an integration constant. We assume that the function $\phi(t)$ decreases during inflation, so $\phi'(N_e)>0$ and, hence, $\omega_0(3M_{Pl}^4-4V_0\xi_0)>0$.

Substituting (\ref{phiN2}) into expression (\ref{VeffC2}) and comparing with expression (\ref{Veffexp}), we obtain:
\begin{equation}
\label{Ceff}
    C_{eff}={}-\frac{3M_{Pl}^4-4V_0\xi_0}{12V_0}\,,
\end{equation}
\begin{equation}
\label{C2}
   C_2={}-\frac{9C_\alpha M_{Pl}^4}{2(3M_{Pl}^4-4V_0\xi_0)}=\frac{3C_\alpha M_{Pl}^4}{8V_0C_{eff}} \,.
\end{equation}
Using the slow-roll equation (\ref{equ00SLR}), we get the parameter
\begin{equation}
\label{eps1min}
    \epsilon_1\simeq\frac{W'}{2W}={}-\frac{C_2}{2\left(N_e+N_0\right)^2}\,.
\end{equation}
The conditions that $0<\epsilon_1<1$ during inflation (for $N_e>0$) and $\epsilon_1=1$ at $N_e=0$ give
\begin{equation}
\label{N_0}
    C_2=-2N_0^2,\qquad C_\alpha=\frac{4N_0^2\left(3M_{Pl}^4-4V_0\xi_0\right)}{9 M_{Pl}^4}\,.
\end{equation}
So, we get that $M_{Pl}^4>4V_0\xi_0/3$ and $\omega_0>0$.

The slow-roll parameter
\begin{equation}
\label{epsilonexp}
\epsilon_2 ={}\frac{2}{N_e+N_0},
\end{equation}
so $\epsilon_2<1$ during inflation if $N_0\geqslant 2$.
If  $8\xi_0V_0<3M_{Pl}^4$, then all slow-roll parameters are less than one during inflation.

Substituting $\phi(N_e)$, given by Eq.~(\ref{phiN2}), into the potential $\tilde{V}(\phi)$ and the function $\tilde{\xi}(\phi)$, we obtain
\begin{equation}
\begin{split}
\label{VxiN}
    \tilde{V}&=V_0\,\exp\left({}-\frac{2N_0^2}{N_e+N_0}\right),\\
    \tilde{\xi}&=\xi_0\,\exp\left(\frac{2N_0^2}{N_e+N_0}\right).
\end{split}
\end{equation}

Using Eqs.~(\ref{ns_C2}) and (\ref{r_cat}), we obtain the inflationary parameters
\begin{eqnarray}\label{ns_N0}
&{}&n_s=1-\frac{2}{N_e+N_0}-\frac{2N_0^2}{\left(N_e+N_0\right)^2},\\
 &{}&r=\frac{16N_0^2\left(3M_{Pl}^4-4V_0\xi_0\right)}{3 M_{Pl}^4\left(N_e+N_0\right)^2}\,.\nonumber
\end{eqnarray}

The observable values of $n_s$, obtained by the telescope Planck~\cite{Planck2018}:
\begin{equation}
n_s=0.965\pm 0.04,
\end{equation}
allows us to restrict values of $N_0$. Indeed, the parameter $N_0$ belongs to the following interval:
\begin{eqnarray}
&2&\leqslant N_0\leqslant 0.0199N_e-0.510\nonumber\\
&{}&{}+0.0102\sqrt{195N_e^2-10000N_e+2500}.\nonumber
\end{eqnarray}
\begin{center}
\begin{figure}[h!tbp]
\includegraphics[width= 7cm, height= 7cm]{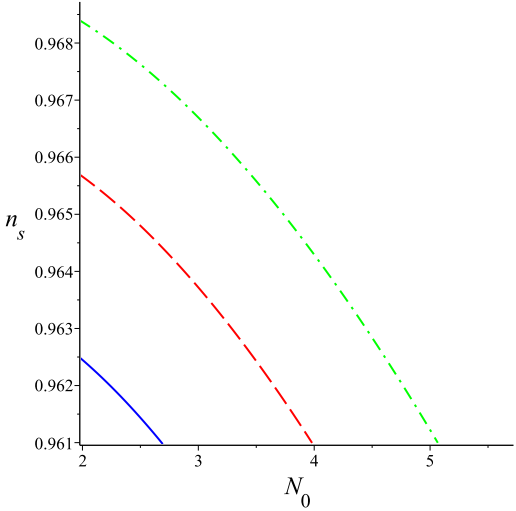}
\caption{The inflationary parameter $n_s$ as a function of $N_0$ for different numbers of e-foldings during inflation: $N_e=55$ (blue solid curve), $N_e=60$ (red dash curve) and $N_e=65$ (green dash-dot curve).
\label{nsN0}}
\end{figure}
\end{center}

We see in Fig.~\ref{nsN0} that for any $55\leqslant N_e\leqslant 65$ it is possible to find suitable values for $N_0$, in particular, the constraint $2\leqslant N_0\leqslant 5.06$ corresponds to $N_e=65$.

The observation condition $A_s=2.1\times10^{-9}$  restricts  values of the model parameters $\xi_0$ and~$V_0$.
From Eqs.~(\ref{As2}) and (\ref{Ceff}) we get
\begin{equation}\label{restriction to xi}
\xi_0=\frac{3M_{Pl}^4}{4V_0} - \frac{(N_b+N_0)^2\,\exp\left({}-\frac{2N_0^2}{(N_b+N_0)}\right)}{32\pi^2A_s N_0^2},
\end{equation}
where $N_b$ is the value of $N_e$, at which the observable value of $A_s$ is calculated. Other restrictions can be obtained from the condition $r<0.065$
(see Ref.~\cite{m} where this model with $\omega_0=2$ has been considered in detail). Note that in the case of $\xi_0=0$, one can get an approximation
the inflationary parameters corresponding to the $R^2$ inflation~\cite{Starobinsky:1980te,Starobinsky:1982,Starobinsky:1983} putting $C_\alpha=1$ and, so, $N_0=\sqrt{3}/{2}\approx 0.87$.

\subsection{Inflationary scenarios with a nontrivial function~$F$}
Let us construct such a model with a nonconstant function $F(\phi)$ that $\phi(N_e)$ and $n_s(N_e)$ are given by Eqs.~(\ref{phiN2}) and (\ref{ns_C2}) correspondingly.
To construct a set of inflationary models with the same function $n_s(N_e)$ we put the condition that $V'_{eff}$ does not change.
It also guarantees that the parameter $A_s$, defined by Eq.~(\ref{As}), does not change.
To get the same function $\phi(N_e)$ in the slow-roll approximation we add the condition that the function $W$ does not change.
 In other words, we consider the model with
\begin{eqnarray}
%\label{FV}
    && F= M_{Pl}^2f(\phi), \nonumber\\
 &&V=f(\phi)\tilde{V}=V_0f(\phi)\exp\left({}-\omega_0\exp\left(-\sqrt{\frac{2}{3C_\alpha}}\frac{\phi}{M_{Pl}}\right)\right),\nonumber
\end{eqnarray}
and
\begin{eqnarray}
\label{Fxi}
\xi(\phi)&=&\left(\xi_0+\frac{3M_{Pl}^4}{4V_0}(f(\phi)-1)\right)\quad\quad\quad\quad\quad\\
&\times&\exp\left(\omega_0\exp\left(-\sqrt{\frac{2}{3C_\alpha}}\frac{\phi}{M_{Pl}}\right)\right),\nonumber
\end{eqnarray}
where $f(\phi)$ is a double differentiable function.

Note that we do not fix the parameter $r(N_e)$:
\begin{equation}
r(N_e)=\frac{12C_\alpha}{f\cdot(N_{e}+N_0)^2}\,,\label{r(r_m)}
\end{equation}
where the parameters $C_\alpha$ and $N_0$ are connected by Eq.~(\ref{N_0}), hence, the observation data~\cite{Planck2018} gives restrictions on the function $f$. Other restrictions on this function can be obtained from the condition that the slow-roll approximation should be satisfied during inflation. We do not change $W(N_e)$, so, the parameters $\epsilon_i$ do not depend on $f$, whereas other slow-roll parameters depend on $f$.

In the following subsections, we consider a few interesting examples of the function $F(\phi)$.

\subsection{The case of an exponential function $F$}
Let us consider the case
\begin{equation}\label{fexp}
    f(\phi)=f_0\exp\left(\beta\omega_0\exp\left(-\sqrt{\frac{2}{3C_\alpha}}\frac{\phi}{M_{Pl}}\right)\right),
\end{equation}
where $\beta$ is a constant. Using Eq.~(\ref{phiN2}),  we get
\begin{equation}\label{Fexp}
    F=M_{Pl}^2f_0\,\exp\left(\frac{2N_0^2\beta}{N_e+N_0}\right),
\end{equation}
and
\begin{equation}
\label{rexp}
    r=\frac{16N_0^2\left(3M_{Pl}^4-4V_0\xi_0\right)}{3 M_{Pl}^4f_0(N_e+N_0)^2}\,\exp\left(-\frac{2N_0^2\beta}{N_e+N_0}\right)\,.
\end{equation}

Also, we obtain
\begin{equation}\label{VNfexp}
    V= f_0V_0\,\exp\left(\frac{2N_0^2(\beta-1)}{N_e+N_0}\right),
\end{equation}
\begin{equation}
\label{xiNfexp}
\begin{split}
    \xi&=\frac{\left(3M_{Pl}^4f_0\,\exp\left(\frac{2\beta N_0^2}{N_e+N_0}\right)-3M_{Pl}^4+4\xi_0 V_0\right)}{4V_0}\\
     &\times \exp\left(\frac{2N_0^2}{N_e+N_0}\right)\,.
\end{split}
\end{equation}

Let us calculate the slow-roll parameters for this model:
\begin{equation}
\label{deltaexp}
   \zeta_1=\frac{2\beta N_0^2}{(N_e+N_0)^2}=2\beta\epsilon_1,\qquad  \zeta_2={}\frac{2}{N_e+N_0}=\epsilon_2,
\end{equation}
\begin{equation*}
%\label{delta1exp}
    \delta_1=\frac{2N_0^2(1+\beta)}{(N_e+N_0)^2}-\frac{2N_0^2(3M_{Pl}^4-4V_0\xi_0)\exp\left(-\frac{2N_0^2\beta}{N_e+N_0}\right)}{3M_{Pl}^4f_0(N_e+N_0)^2}\,.
\end{equation*}
\begin{eqnarray*}
%\label{delta2exp}
  &&  \delta_2=\frac{2}{N_e+N_0}\\
&&{}+\frac{2(3M_{Pl}^4-4\xi_0V_0)N_0^2\beta\,\exp\left(-\frac{2N_0^2\beta}{N_e+N_0}\right)(N_e+N_0)^{-2}}
    {\left[3 M_{Pl}^4f_0(1+\beta)-(3M_{Pl}^4-4\xi_0V_0)\,\exp\left(-\frac{2N_0^2\beta}{N_e+N_0}\right)\right]}.
\end{eqnarray*}
The condition $|\beta|\leqslant 1/2$ is necessary to get  $|\zeta_1|<1$ during inflation.

To fix $f_0$ we put the following condition at the end of inflation:
\begin{equation*}
F(\phi(0))=M_{Pl}^2,
\end{equation*}
 therefore,
\begin{equation}
f_0=\exp\left(-2N_0\beta\right)\,.
\end{equation}

To get the slow-roll evolution during inflation we restrict the value of the product $\xi_0V_0$, after this the parameters $\xi_0$ and $V_0$ can be obtained using Eq.~(\ref{restriction to xi}).

At $N_e=0$, we get
\begin{equation*}
\delta_1(0)=\frac{8J}{3}+2\beta, \qquad \delta_2(0)=\frac{2}{N_0}-\frac{2\beta(4J-3)}{4J+3\beta},
\end{equation*}
where $J\equiv V_0\xi_0/M_{Pl}^4$.

\begin{figure}[h!tbp]
\includegraphics[width= 7cm, height= 7cm]{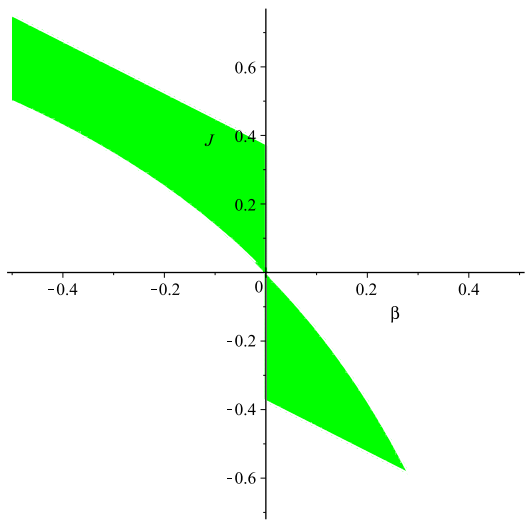}
\caption{Possible values of parameters $J$ and $\beta$ are in green domain. \label{inequJbeta}}
\end{figure}

Let us consider the case $N_0=2$ in detail.
We get
\begin{equation}\label{conddelta1}
    {}-\frac{1}{2}\leqslant \frac{4}{3}J+3\beta \leqslant \frac{1}{2}\,,
\end{equation}
\begin{equation}\label{conddelta2}
    {}-2\leqslant \frac{2\beta(3-4J)}{4J+3\beta} \leqslant 0.
\end{equation}
Also, we have the conditions $|\beta|\leqslant 1/2$. So,  it follows from inequalities (\ref{conddelta1}) that $|J|\leqslant 3/4$. Note that $J=3/4$ is excluded (see Eq.~(\ref{VeffC2}) for the effective potential). In Fig.~\ref{inequJbeta}, the green domain corresponds to the values of parameters $J$ and $\beta$ that satisfy inequalities (\ref{conddelta1}) and (\ref{conddelta2}). At $\beta=0$, we get the initial model with a constant $F$.

Substituting the chosen values of the constants into formulae (\ref{As2}) and (\ref{rexp}), we obtain
\begin{equation}
A_s=\frac{V_0(N_b+2)^2}{32\pi^2M_{Pl}^4(3-4J)}\exp\left(-\frac{8}{N_b+2}\right)\,,
\end{equation}
\begin{equation}
\label{rN02}
    r={}\frac{64(3-4J)}{3(N_b+2)^2}\exp\left(\frac{4\beta N_b}{N_b+2}\right)\,.
\end{equation}

The values of the inflationary parameter $r$ and the corresponding values of $V_0$ and $\xi_0$ for $N_b=60$ are presented in Table~\ref{T1}.
For any values of these parameters,  $n_s=928/961\simeq 0.96566$ and $A_s=2.1\cdot10^{-9}$.
One can see that the parameter $r$ increases with growth of $J$ and all values of $r$, but one, do not contradict the observation data.\\

\begin{table}[h]
\begin{center}
\caption{Model parameters and the corresponding values of $r$ for the exponential function~$F$.}
\begin{tabular}{|c|c|c|c|c|}
    \hline
     $\beta$ & $J$ & $V_0/{M_{Pl}^4}^{\phantom{7}}$ & $\xi_0$ & $r$ \\
      \hline
     $-0.5$ & $0.72$ & $ 2.3556\cdot{10^{-11}}^{\phantom{7}}$ & $3.0565\cdot 10^{10}$ & $0.0001$ \\
      \hline
     $-0.5$ & $0.5 $ & $1.9630\cdot{10^{-10}}^{\phantom{7}}$ & $2.5471\cdot 10^9 $  &  $0.0008$\\
     \hline
      $-0.3$& $0.5$ & $1.9630\cdot{10^{-10}}^{\phantom{7}}$ & $2.5471\cdot 10^9$ & $0.0017$ \\
     \hline
      $-0.1$& $0.45$ & $2.3556\cdot{10^{-10}}^{\phantom{7}}$ & $1.9103 \cdot 10^9$ & $0.0045$ \\
      \hline
      $-0.1$& $0.2$ & $4.31863\cdot{10^{-10}}^{\phantom{7}}$ &  $4.6311\cdot 10^8 $& $0.0083$ \\
      \hline
      $0$& $0.2$ & $4.3186\cdot{10^{-10}}^{\phantom{7}}$ & $4.6311\cdot 10^8 $ & $0.0122$ \\
      \hline
      $0.1$& $-0.2$ & $7.4595\cdot{10^{-10}}^{\phantom{7}}$ & $-2.6812\cdot 10^8 $ & $0.0311$ \\
      \hline
      $0.1$& $-0.4$ & $9.0299\cdot{10^{-10}}^{\phantom{7}}$ & $-4.4297\cdot 10^8 $ & $0.0376$ \\
      \hline
      $0.2$& $-0.4$ & $9.0299\cdot{10^{-10}}^{\phantom{7}}$ & $-4.4297\cdot 10^8 $ & $0.0554$ \\
    \hline
      $0.25$& $-0.45$ & $9.4225\cdot{10^{-10}}^{\phantom{7}}$ & $-4.7758\cdot 10^8$ & $0.0701$ \\
    \hline
  \end{tabular}
   \label{T1}
\end{center}
\end{table}

\subsection{The coupling function $F$ with a constant term}

Let us consider another form of nonminimal couplings that tends to a constant at small values of the field:
\begin{equation}
F=M^2_{Pl} f(\phi), \qquad f(\phi)=\frac{1+\tilde{f}{(\phi)}}{1+\tilde{f}(\phi_{end})},
\end{equation}
where
\begin{equation}
\tilde{f}(N_e)= f_0\exp\left(-\frac{\omega_0}{2}\exp\left(\sqrt{\frac{2}{3C_\alpha}}\frac{\phi}{M_{Pl}}\right)\right),\label{tilde(f)(phi)}
\end{equation}
$f_0$ is a positive constant and $\phi_{end}=\phi(0)$ is the value of $\phi$ at the end of inflation.

To construct a model leading to an appropriate inflationary scenario it is convenient to use the e-folding number formulation.
Using Eq.~(\ref{phiN2}), we get
\begin{equation}
\tilde{f}(N_e)=f_0\exp\left(-\frac{N_0^2}{N_e+N_0}\right)\label{tilde(f)}\,.
\end{equation}
With the help of Eqs.~\eqref{VxiN}, \eqref{Fxi}, and \eqref{tilde(f)}, we obtain the functions $\xi$ and $V$ in the following form:
\begin{eqnarray}
 \xi&=&\left[ {\xi_0}+\frac{3{M}^{4}_{Pl}}{4{{V_0}}} \left(\frac{ 1+{ f_0}\,{\exp\left(-
{\frac {{N_0^2}}{N_e+{N_0}}}\right)} }{ 1+{ f_0}\,\exp
\left(-\frac {{N_0^2}}{{N_0}}\right)}-1 \right) \right]\nonumber\\&\times&{\exp\left(\frac{2N_0^2}{N_e+N_0}\right)},  \\
 V&=&\frac{V_0\left[ 1+{ f_0}\,{\exp\left(-{\frac {{N_0^2}}{N_e+{N_0}}}\right)}
 \right]{\exp\left(\frac{2N_0^2}{\left( N_e+N_0\right)}\right)}}{ 1+{f_0}\,{\exp\left(-N_0\right)}} \,.
\end{eqnarray}
For the considering model the slow-roll parameters $\epsilon_1$ and $\epsilon_2$ are coincide  with \eqref{epsilonexp}, the slow-roll parameters  $\zeta_1$ and $\zeta_2$ have the following form:
\begin{eqnarray}
  &&\zeta_1=-\frac{f_0N_0^2\exp\left(-{\frac {{N_0^2}}{N_e+{N_0}}}\right)}{ \left( N_e+N_0 \right)^{2}\left( 1+f_0\exp\left(-{
\frac {{N_0^2}}{N_e+{N_0}}}\right)\right)},\nonumber  \\
%\label{zeta1}
  &&\zeta_2=\frac{2}{N_e+N_0}-\frac{N_0^2\exp\left(\frac{N_0^2}{N_e+N_0}\right)}{\left( N_e+N_0\right)^{2}\left(\exp\left(\frac{{N_0^2}}{N_e+{N_0}}\right)+f_0\right)}\,.\nonumber
  %\label{zeta2}
  \end{eqnarray}

We consider $f_0>0$, hence,  $\zeta_1<0$. Also, we see that $-1<\zeta_1<0$ and $0<\zeta_2<1$ during inflation for $N_0\geqslant 2$.

  To simplify expression of the slow-roll parameters $\delta_1$ and $\delta_2$ we  introduce a new constant $K$, namely we do redesignation of the constant $V_0$:
\begin{equation*}
{V_0}=\frac{A_s{M}^4_{Pl} \left( 24{\pi}^{2}N_0^{2}+K\right)}{\left(N_b+N_0\right)^{2}\left(1+f_0\exp(-N_0)\right)\exp\left(-\frac{2N_0^2}{N_b+N_0}\right) }\,.
\end{equation*}

After such supposition, the slow-roll parameters $\delta_1$ and $\delta_2$   can be presented in the following form:
\begin{eqnarray}
\delta_1&=&{\frac { 12\,
{\pi}^{2}{ f_0}\,{{N_0}}^{2}{\exp\left(-{\frac{N_0^{2}}{N_e+{N_0}}}\right)}-K }
{ 12{\pi}^{2}\left( N_e+{N_0} \right)^{2}\left( 1+{ f_0}\,\exp\left(-{\frac{N_0^{2}}{N_e+{N_0}}}\right) \right)}}\,,
%\label{d1K}
\nonumber \\
\delta_2&=&\frac{2K}{(N_e+N_0)\left(K-12\pi^2f_0N_0^2\exp\left({}-
\frac{N_0^2}{N_e+N_0}\right)\right)}\nonumber
\end{eqnarray}
$+\frac{N_0^2\left( 12\,{\pi}^{2}N_0^{2}+K-24\pi^2(N_e+N_0)\right)f_0\exp\left(-{\frac{N_0^2}{N_e+N_0}}\right)}{\left(1+f_0\exp\left(-\frac{N_0^2}{N_e+N_0}\right)\right)( N_e+N_0)^{2}\left[K-12\pi^2f_0N_0^2\exp\left(-
\frac{N_0^2}{N_e+N_0}\right) \right]}
$\\
$-\frac{24\,{\pi}^{2}N_0^{2}f_0^{2} \left( N_e+{N_0} \right)\exp\left(-{\frac{2N_0^{2}}{N_e+N_0}}
\right)}{\left(1+f_0\exp\left(-\frac{N_0^2}{N_e+N_0}\right)\right)( N_e+N_0)^{2}\left[K-12\pi^2f_0N_0^2\exp\left(-
\frac{N_0^2}{N_e+N_0}\right) \right]}.$\\

The slow-roll parameters $\delta_1$ and $\delta_2$ at $N_e=0$ are as follows:
\begin{equation*}
\delta_1={\frac {12\,{\pi }^{2}{f_0}\,{{N_0}}^{2}{\exp(-{N_0})}-K}{12\left( 1+{f_0}\,{\exp(-{N_0})} \right) {{N_0}}^{2}{\pi }^{2}}}
\end{equation*}
\begin{equation*}
  \delta_2=\frac{2}{N_0}+{\frac {f_0\exp(-N_0)\left( 12\pi^2N_0^2+K\right) }{ \left( 1+f_0\exp(-N_0)\right) \left[K-12\pi^2f_0N_0^2\exp(-N_0)\right] }}
\end{equation*}

Solving inequalities $|\delta_1|\leqslant 1$ and  $|\delta_2|\leqslant 1$  at $N_e=0$, we get the area of appropriate values of parameters $f_0$ and $K$  restricted by the following curvatures:
\begin{enumerate}
  \item the lower line is  $K=K_l=-12\pi^2N_0^2$ (corresponds to $\delta_1=\delta_2=1$ at $N_e=0$)
  \item the upper line is \\$K=K_u=\frac{12{\pi }^{2}f_0N_0^{2}{\exp(-N_0)}\left( 2+f_0 \left( N_0+2 \right)\exp(-N_0)\right)}{2\,f_0\left(N_0+1\right){\exp(-N_0)}+N_0+2},$ \\ (corresponds to $\delta_2=-1$ at $N_e=0$)
\end{enumerate}

The tensor-to-scalar ratio $r$ can be presented in the following form:
\begin{equation*}
r=\frac{2\left( 24{\pi}^{2}N_0^{2}+K \right)}{3{\pi}^{2} \left( 1+{f_0}\,{\exp\left(-{\frac{N_0^2}{N_e+N_0}}\right)} \right) \left(N_e+N_0\right)^{2}}.
\end{equation*}

 In the case of the lower boundary of area appropriate values of parameters $0< f_0<\infty$ and $K=-12\pi^2N_0^2$, the expressions of the slow-roll parameters can be simplified:
\begin{equation*}
 \delta_1=\epsilon_1=\frac{N_0^2}{(N_e+N_0)^2},\quad  \delta_2=\epsilon_2=\frac{2}{N_e+N_0}
\end{equation*}

Also we get a simplification of  $V_0$:
\begin{equation*}
V_0=\frac{12A_s{M}^4_{Pl}{\pi}^{2}N_0^{2}}{\exp\left(-\frac{2N_0^{2}}{N_b+N_0}\right)\left( 1+f_0\exp\left(
-N_0\right)\right)(N_b+N_0)^{2}}
\end{equation*}
and the tensor-to-scalar ratio
\begin{equation*}
r=\frac{8\,{{N_0}}^{2}}{ \left( 1+{ f_0}\,{\exp\left(-{\frac {{{N_0}}^{2
}}{N_e+{N_0}}}\right)} \right)\left( N_e+{N_0} \right)^{2}}\,.
\end{equation*}

From the previews analysis, it is evidently that the $N_0=2$ is an appropriate value. Here we should note that expressions for parameters  $\delta_1$ and $\delta_2$ are coincide  with expressions for parameters  $\epsilon_1$ and $\epsilon_2$. At $f_0=0$, parameters  $\zeta_1$ and $\zeta_2$  disappear because $F$ becomes a constant.

\begin{figure}[h!tbp]
\includegraphics[width= 7cm, height= 7cm]{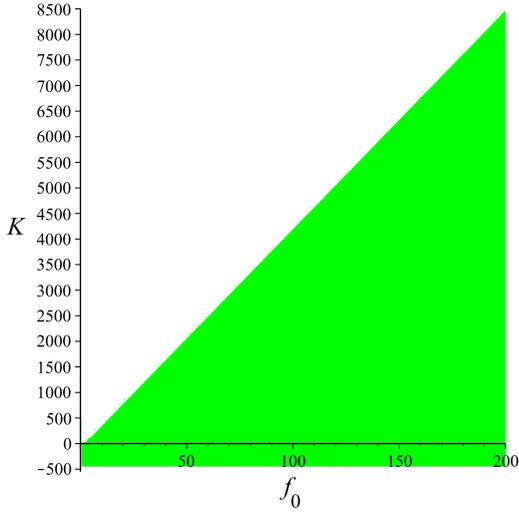}
\caption{The domain of appropriate values of parameters $f_0$ and $K$ is green colored. \label{K(f0)}}
\end{figure}

Considering conditions $|\delta_1|\leqslant 1$ and  $|\delta_2|\leqslant 1$ at $N_e=0$ and $N_0=2$, we get the following inequalities:
\begin{eqnarray}
             -1&\leqslant  & {\frac {48\,{\pi}^{2} f_0\,{\exp(-2)}-K}{48 \left( 1+{
 f_0}\,{\exp(-2)} \right) {\pi}^{2}}}\leqslant  1 \nonumber \\
             -1 &\leqslant & {\frac{-48\,{\pi}^{2}f_0^2\,{\exp(-4)}+2\,K{ f_0}\,{
\exp(-2)}+K}{ \left( 1+{ f_0}\,{\exp(-2)} \right) \left( -
48\,{\pi }^{2}{ f_0}\,{\exp(-2)}+K \right) }}\leqslant  1 \nonumber
\end{eqnarray}
Solving these  inequalities, we get the area restricted by the following curvatures:
\begin{enumerate}
  \item the lower line is  $K=K_l=-48\pi^2$ (corresponds to $\delta_1=\delta_2=1$ at $N_e=0$)
  \item the upper line is $$K=K_u={\frac{48\,{\pi }^{2}{ f_0}\,{\exp(-2)} \left( 2\,{ f_0}\,{
\exp(-2)}+1 \right) }{3\,{ f_0}\,{\exp(-2)}+2}}$$ (corresponds to $\delta_2=-1$ at $N_e=0$)
\end{enumerate}
 The corresponding area is green colored  in  Fig.~\ref{K(f0)}.

Numerical analysis of the obtained model with $K=K_{l}$, $N_0=2$, and $N_b=60$ shows that  $n_s =0.965$ and $r<0.065$ at any positive $f_0$ and the slow-roll regime is satisfied during inflation.
We present the values of  $V_0$, $\xi_0$, $C_\alpha$, $r$ for $N_0=2$, $N_b=60$, $A_s=2.1\cdot 10^{-9}$, and at several values of $f_0$: $0\leqslant  f_0\leqslant 8$ in Table~\ref{T2}.\\

\begin{table}[h]
\begin{center}
\caption{Model parameters and the corresponding values of $r$ for the function $F$ with a constant term in the case of $K=K_l$.}
  \begin{tabular}{|c|c|c|c|c|}
    \hline
     $ f_0$ & $V_0/{M^4_{Pl}}^{\phantom{7}}$ & $\xi_0$ & $C_\alpha$ & $r$ \\
      \hline
     $0$ & $2.9446\cdot{10^{-10}}^{\phantom{7}}$ & $1.2736\cdot 10^9$ & $2.6667$ & $0.0083\,$ \\
      \hline
     $1$ & $2.5937\cdot{10^{-10}}^{\phantom{7}}$ & $1.6183\cdot 10^9$ & $2.3488$  &  $0.0044\,$\\
     \hline
      $2$& $2.3173\cdot{10^{-10}}^{\phantom{7}}$ & $1.9631\cdot 10^9$ & $2.0986$ & $0.0028\,$ \\
      \hline
      $3$& $2.0943\cdot{10^{-10}}^{\phantom{7}}$ & $2.3078\cdot 10^9$ &  $1.8966$& $0.0022 \,$ \\
      \hline
      $4$& $1.9103\cdot{10^{-10}}^{\phantom{7}}$ & $2.6524\cdot 10^9$ & $1.7301$ & $0.0017\,$ \\
      \hline
      $5$& $1.7562\cdot{10^{-10}}^{\phantom{7}}$ & $2.9972\cdot 10^9$ & $1.5904$ & $0.0015\, $ \\
      \hline
      $6$& $1.6251\cdot{10^{-10}}^{\phantom{7}}$ & $3.3420\cdot 10^9$ & $1.4717$ & $0.0012 \,$ \\
      \hline
      $7$& $1.5121\cdot{10^{-10}}^{\phantom{7}}$ & $3.6868\cdot 10^9$ & $1.3694$ & $0.0011\,$ \\
    \hline
      $8$ & $1.4138\cdot{10^{-10}}^{\phantom{7}}$ & $4.0313\cdot 10^9$ & $1.2804$ & $0.0010\,$ \\
    \hline
  \end{tabular}
  \label{T2}
\end{center}
\end{table}

\begin{figure}[h!tbp]
\includegraphics[width= 7cm, height= 7cm]{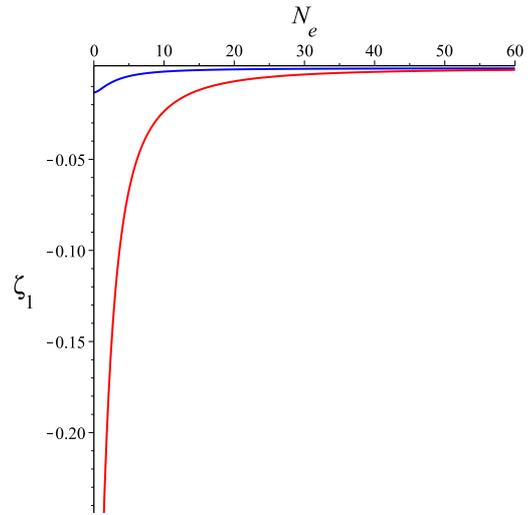}\quad\quad\includegraphics[width= 7cm, height= 7cm]{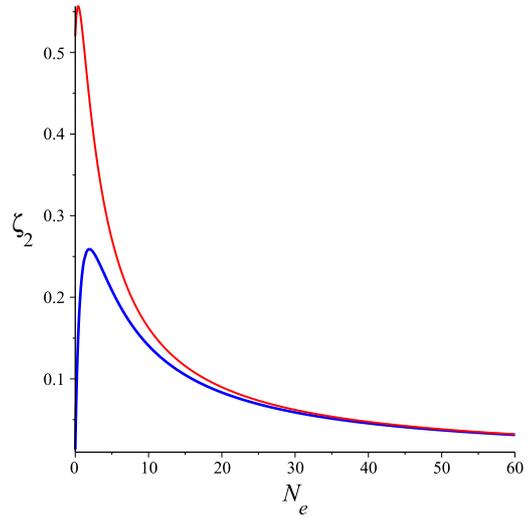}$$ $$
\caption{The parameters $\zeta_1$ and $\zeta_2$ as functions of $N_e$. The blue lines correspond to $f_0=0.1$,  the red lines  correspond to  $f_0=8$. \label{zeta}}
\end{figure}

To clarity behavior of slow-roll parameters we present graphics of $\zeta_1$ and $\zeta_2$ at $ f_0=0.1$ and  $ f_0=8$  in  Fig.~\ref{zeta}.\\

In the case of $K=K_u$, the expressions of the slow-roll parameters $\delta_1$ and $\delta_2$ for an arbitrary $N_0$ are rather long and we present only $\delta_1$ at $N_0=2$:\\
$
\delta_1=\frac{4f_0\left[3f_0\exp\left(-\frac{2(4+N_e)}{N_e+2}\right)+2\exp\left(-\frac{4}{ N_e+2}\right)-2f_0\exp(-4)-\exp(-2)\right] }{ \left( 3\,{ f_0}\,{\exp(-2)}+2 \right)\left( 1+{ f_0}\,{\exp\left(-\frac{4}{ N_e+2}\right)}\right) \left( N_e+2 \right)^{2}}
$

\begin{figure}[h!tbp]
\includegraphics[width= 7cm, height= 7cm]{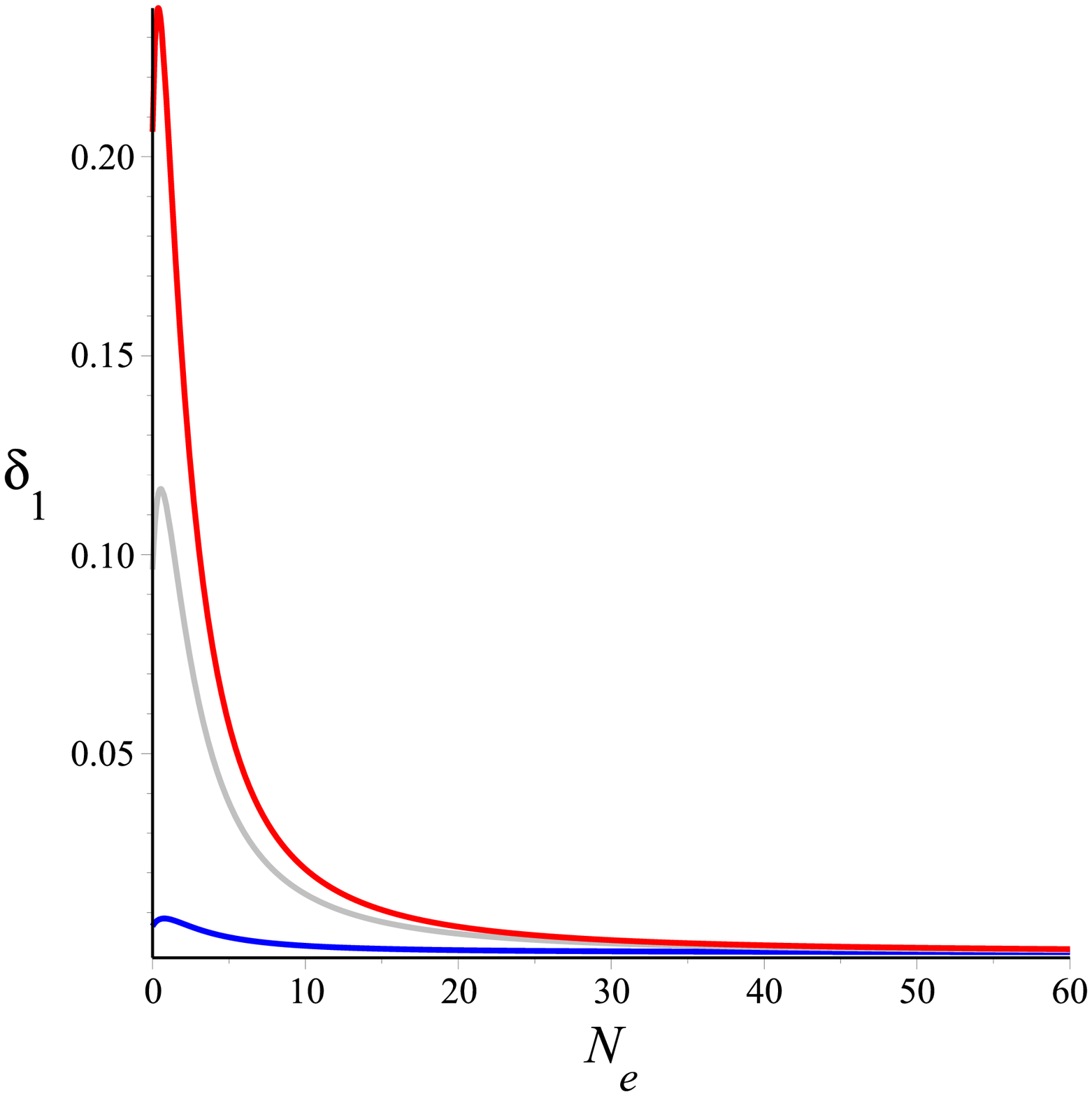}\quad\quad\includegraphics[width= 7cm, height= 7cm]{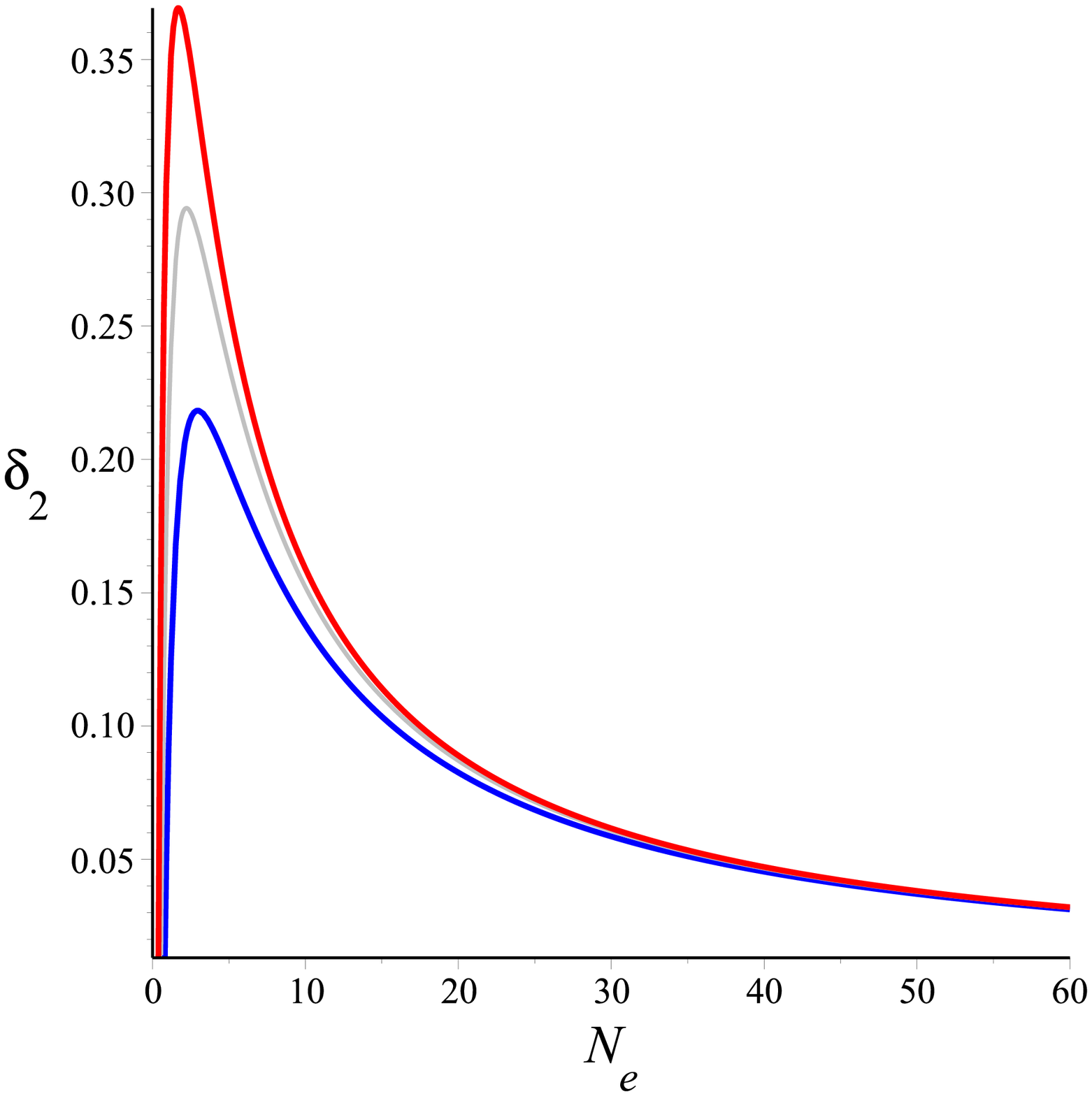}$$ $$ $$ $$
\caption{The blue lines correspond to $ f_0=0.1$, the grey lines correspond to  $ f_0=2$, and the red lines  correspond to  $f_0=8$. The parameters $\delta_1$ and $\delta_2$ are presented  at $K=K_u$. \label{delta12}}
\end{figure}

To clarity behavior of slow-roll parameters at $K=K_u$ we present graphics of $\delta_1$ and $\delta_2$ at  $f_0=0.1$,   $ f_0=2$,  and  $ f_0=8$ in  Fig.~\ref{delta12}.\\

Let us numerically analyze the obtained models with $K=K_u$. If  $N_0=2$ and $N_b=60$, then $n_s =0.965$ and $r<0.065$ at any $f_0\geqslant 0$ and the slow-roll regime is satisfied during inflation.

We present the values of   $V_0$,  $C_\alpha$, $r$ at $N_0=2$, $N_b=60$, $A_s=2.1\cdot 10^{-9}$  and at several values of $ f_0$: $0.001\leqslant  f_0\leqslant 8$ in Table~\ref{T3}.\\

\begin{table}[h]
\begin{center}
\caption{Model parameters and the corresponding values of $r$ for the function $F$ with a constant term in the case of $K=K_u$.}
  \begin{tabular}{|c|c|c|c|c|}
    \hline
     $ f_0$ & $V_0/M^4_{Pl}$ & $\xi_0$& $C_\alpha$ & $r$ \\
      \hline
     $0.001$ & $\,\,5.8887\cdot{10^{-10}}^{\phantom{7}} $ & $ \,\,1.294\cdot10^5\,$ & $\,\,5.3334\,$ & $\,\,0.017$ \\
      \hline
     $1$ & $\,\,5.3727\cdot{10^{-10}}^{\phantom{7}}$ & $ \,\,1.224\cdot10^8\,$ & $\,\,4.8660\, $  &  $\,\,0.0091 $\\
     \hline
      $2$& $\,\,4.9784\cdot{10^{-10}}^{\phantom{7}}$ & $\,\, 2.329\cdot10^8\,$ & $\,\,4.5089\,$ & $\,\,0.0063 $ \\
      \hline
      $3$& $\,\,4.6674\cdot{10^{-10}}^{\phantom{7}}$ & $ \,\,3.334\cdot10^8\,$ &  $\,\,4.2272\, $& $\,\,0.0048$ \\
      \hline
      $4$& $\,\,4.4148\cdot{10^{-10}}^{\phantom{7}}$ & $\,\, 4.253\cdot10^8\,$ & $\,\,3.9985\,$ & $\,\,0.0040$ \\
      \hline
      $5$& $\,\,4.2062\cdot{10^{-10}}^{\phantom{7}}$ & $ \,\,5.096\cdot10^8\,$ & $\,\,3.8095\, $ & $\,\,0.0035$ \\
      \hline
      $6$& $ \,\,4.0305\cdot{10^{-10}}^{\phantom{7}}$ & $ \,\,5.872\cdot10^8\,$ & $\,\,3.6505\,$ & $\,\,0.0030$ \\
      \hline
      $7$& $ \,\,3.8802\cdot{10^{-10}}^{\phantom{7}}$ & $\,\, 6.592\cdot10^8\,$ & $\,\,3.5143\,$ & $\,\,0.0030$ \\
    \hline
      $8$ & $\,\, 3.7508\cdot{10^{-10}}^{\phantom{7}}$ & $\,\, 7.261\cdot10^8\,$ & $\,\,3.3971\,$ & $\,\,0.0026$ \\
    \hline
  \end{tabular}
  \label{T3}
\end{center}
\end{table}

In the case of $f_0=0$, when the model coincides with the minimal coupling model (see Section 4.1), the slow-roll parameter $\delta_2=\frac{2}{N_e+N_0}$  does not depend on $K$.
In this case, the restriction to $K$ can be obtained from the consideration of $\delta_1$. The condition
 \begin{equation*}
 -1\leqslant\delta_1={}-\frac{K}{12\pi^2(N_e+N_0)^2}
\end{equation*}
should be satisfied and we get  $K={K_u^0}=12\pi^2\,N_0^2$. So, to satisfy the slow-roll regime we should choose $|K|\leq 12 \pi^2 N_0^2$. The numeral estimation of model parameters at $f_0=0$, $K=K_u^0$, $N_0=2$, and $N_b=60$ gives the following values: $V_0/M^4_{Pl}=8.8336\cdot10^{-10}$, $\xi_0=-4.2452\cdot10^8$, and $C_\alpha=8.0001$. The corresponding value of the tensor-to-scalar ratio is $r=0.0250$. Obviously, at $f_0=0$, the slow-roll parameters  $\zeta_i$ are disappear due to $F'=0$.

In the limit $f_0 \rightarrow +\infty$, we obtain $V_0/M^4_{Pl}\approx1.963\cdot10^{-10}$, $\xi_0\approx 2.5471\cdot10^9$, $ C_\alpha\approx 1.7778$, and $r\approx0.0008$.

\section{Conclusion}
In this paper, inflationary scenarios of the Einstein-Gauss-Bonnet gravity have been considered.
We focus on inflationary models with a scalar field nonminimally coupled both with the Ricci curvature scalar
and with the Gauss-Bonnet term. To construct viable inflationary scenarios that do not contradict the observation data we consider the inflationary parameters as functions of e-foldings.

The application of the slow-roll regime to the model allows to present the scalar spectral index $n_s(N_e)$ and the amplitude of the scalar perturbations $A_s(N_e)$
in terms of derivatives of the effective potential. The main idea of the proposed method is the construction of a set of inflationary scenarios with one and the same effective potential. It allows us to construct new models with suitable values of $n_s$ and $A_s$. Starting from an inflationary model with the Gauss-Bonnet term and a constant $F$ that does not contradict observation data, one can construct models with a nonconstant function $F$ with the values of parameters $n_s$ and $A_s$, but with different values of $r$. In other words, to compare the model predictions with observation data it is sufficient to check the tensor-to-scalar ratio in the nonminimal coupling model: $\tilde{r}/f(N_b)<0.065$, where $\tilde{r}$ is the tensor-to-scalar ratio of the known minimally coupled model~\cite{m}. Also, we treat with care to the slow-roll regime during inflation that allows us to restrict the free parameters of the models obtained.

In distinguish to the cosmological attractor approach and the method proposed in Ref.~\cite{Odintsov:2018zhw}, we do not fix $r(N_e)$, but fix $\phi(N_e)$ and $n_s(N_e)$. We also use the known inflationary scenario with a constant function $F=M^2_{Pl}$ propose in Ref.~\cite{m} to construct sets of inflationary models with nonconstant functions $F$ that is equal to $M^2_{Pl}$ at the end of inflation.

In this paper, we restrict ourselves to inflationary models for which all slow-roll parameters are small during inflation. At the same time, it would be interesting to consider models with a constant
positive potential and with the potential $V=CF^2$ that do not satisfy this restriction. We plan to do this in future using numerical analysis of evolution equations without the slow-roll approximation.

{ \ }

\textbf{{Acknowledgements}}

This work is partially supported by  the Russian Foundation for Basic Research grant No. 20-02-00411.

\end{document}